\documentclass[11pt]{article}
\usepackage{geometry}                % See geometry.pdf to learn the layout options. There are lots.
\usepackage{graphicx}
\usepackage{amssymb}
\usepackage{epstopdf}

\usepackage{amsmath, amssymb, graphics}

\newcommand{\mathsym}[1]{{}}

\title{Energy Calibration of Underground Neutrino Detectors using a ~100 MeV electron accelerator.}
\author{Sebastian~White and Vitaly~Yakimenko \\
Brookhaven National Laboratory\\
Upton, NY 11973 USA}

\begin{document}
\maketitle
\section{Introduction}

	An electron accelerator  in the 100 MeV range, similar to the one used at BNL's Accelerator test Facility(Table 1), for example, 
would have some advantages as a calibration tool for water cerenkov or Liquid Argon neutrino detectors. Alternatives like Michel electrons from cosmic ray muons
and the low energy (5-16 MeV) linac used by Super Kamiokande don't cover the full energy range needed for the future long baseline program and have poor
time definition.

	In LBNE the electron neutrino energy distribution will be fit to
obtain the CP violation parameter. One of the contributions to
the energy scale (relative and absolute) uncertainty is the electron
resolution function.  An accelerator that allows for a scan of the
electron energy over the range of 0.1 to a few GeV would be very important
for this measurement.
	Our goal is to address absolute energy scale calibration up to an electromagnetic shower energy of at least 1 GeV.

	Super Kamiokande's linac intensity was reduced from $10^6$ electron/pulse to ~1 electron/pulse using constraints on the beam divergence and 
	momentum spread. The pulse width was 1-2 $\mu$sec.

	Here we investigate large angle Mott scattering as a technique to reduce the beam intensity to $ \sim$1/2 electron/pulse. Since it is difficult
	to operate electron accelerators at $\leq10^7$ e/pulse we need this level of reduction. This is also potentially an
	attractive way to produce a benchtop secondary beam. 
	
		It would seem that this scheme could only work if the primary beam can be brought close to 
	the detector. However a focusing triplet with a length of 0.5 meters and 2 centimeter aperture could be placed at 0.5 to 1 meter from the
	scattering target thereby extending this distance by at least a factor of 10. 
		
	In this note we calculate the thickness of targets that would be needed to produce a scattered beam of 1/2 electron/pulse and the amount of emittance growth they would produce in the beam downstream of the target. If the multiple scattering is small enough it should be possible to have many secondary beam
	ports along the length of the accelerator. 
	
	Surprisingly we find that the beam emittance blowup caused by the scattering target doesn't depend very much on the choice of material in spite of the very strong contribution of the form factor. So the conclusion of our optimization exercise is that there isn't really an optimum choice of target. But lighter targets are better than gold.
	
	This technique could provide several energy calibration points. At large angles, depending on the target material, well
	resolved inelastic peaks with a typical ~4 MeV spacing, corresponding to the nuclear level structure, appear in the scattered electron energy spectrum, as seen in Figure~\ref{fig:peaks}.
	The energy resolution of the Icarus\cite{Icarus} detector is roughly $4\%$ at 100 MeV. A water Cerenkov detector with 25$\%$ coverage would have a $\sim 4.5\%$ stochastic term at this energy.
	
	With this technique one could also take advantage of the strict proportionality between the accelerator intensity and the scattered electron multiplicity in a pulse.
	The beam intensity in the ATF, for example, can easily be adjusted over a range of more than x10 and it is well measured so it would be straightforward to vary the mean energy deposit within a pulse shorter than 1 nsec  from the beam energy to $\sim$10 times the beam energy. Icarus\cite{Icarus} shows that there are different problems
	in analyzing high energy showers compared to low energy showers but multiple electron showers have also been used in calibrations- for example at the Final focus testbeam.
	
\begin{figure}[bt]
\begin{center}
\raisebox{1cm}{
\includegraphics[width=75mm]{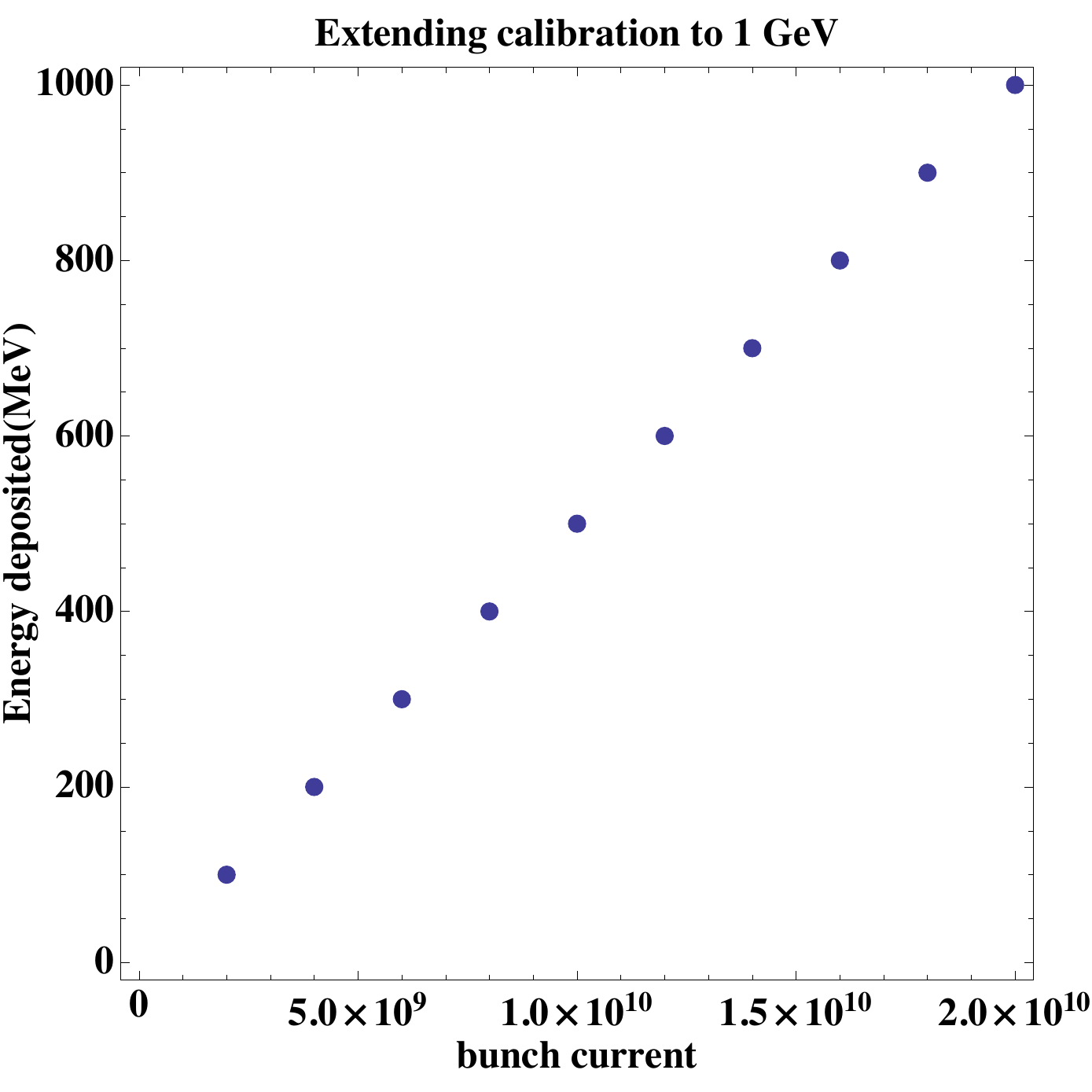}
\includegraphics[width=75mm]{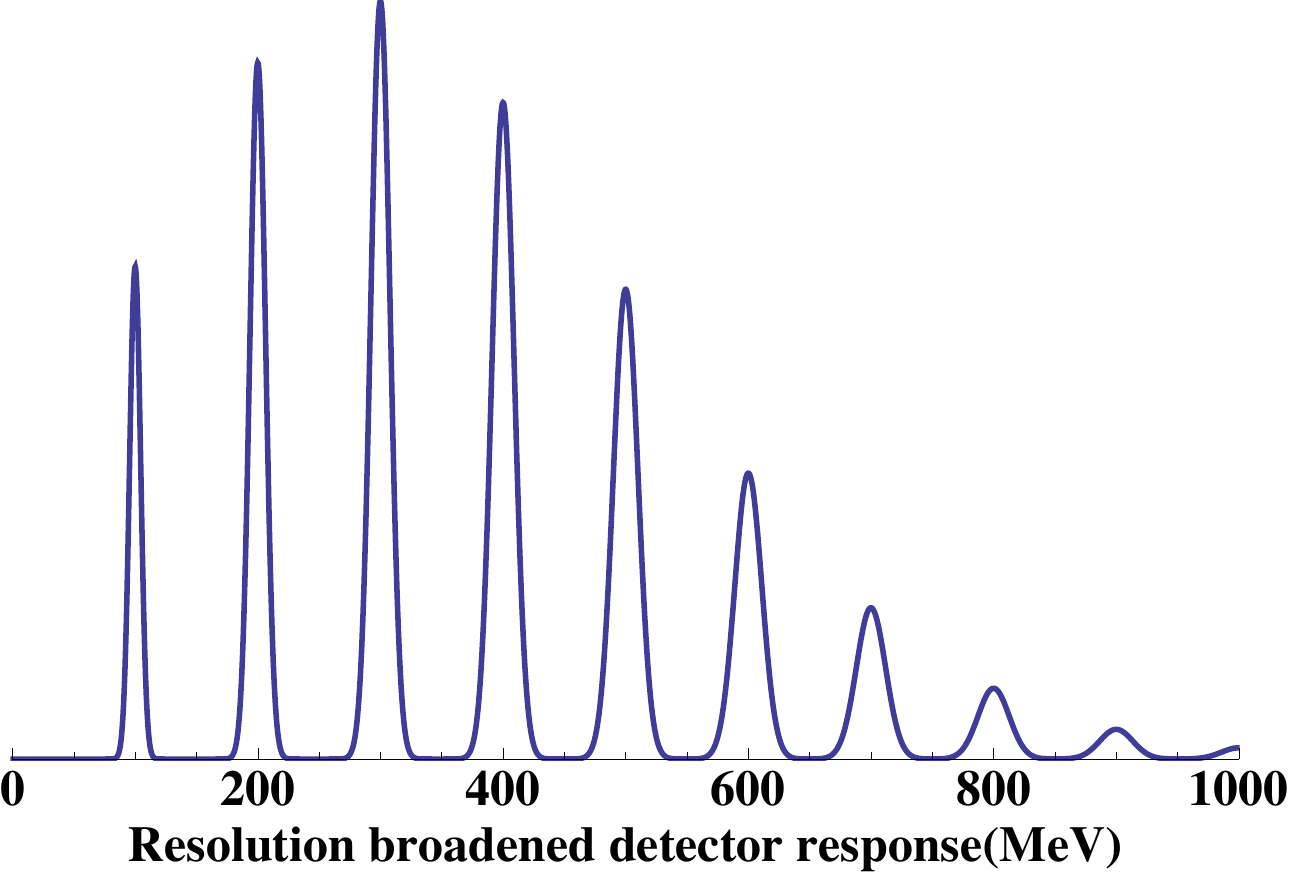}
}
\caption[  The scattered electron rate is linear in the primary beam intensity and this is well measured. So it will be  straightforward to adjust
the current to get an average, an energy deposit of several times $E_{beam}$.]{\label{fig:linear}
The scattered electron rate is linear in the primary beam intensity and this is well measured. So it will be  straightforward to adjust
the current to get, on average, an energy deposit of several times $E_{beam}$. The right plot shows schematically the expected energy response
for a 100 MeV beam with intensity of  8*$10^9$ electrons/pulse.  The peaks are smeared assuming an energy resolution corresponding to photostatistics of 5 photoelectrons/MeV
expected for a water Cerenkov detector with 25$\%$ coverage. The stochastic term for LAr obtained by Icarus\cite{Icarus} is somewhat better- ie $\sim0.33/\sqrt{E(MeV)}$.
}
\end{center}
\end{figure}

	For some purposes it could be useful that, because of statistical fluctuations in the electron multiplicity, one can simultaneously measure the response in
	several energy peaks as illustrated in fig.~\ref{fig:linear}.
	
		We will return to the question of construction and operating costs.
	
	\begin{table}[ht]
\caption{An example: ATF parameters}
\centering
\begin{tabular}{c c}
  \hline
 Beam Energy &  80 MeV \\
Bunch Intensity &  $10^8-10^{10}$ e/pulse \\
Rep Rate &  1 Hz \\
Bunch Length &  3 picoseconds \\
Length &  60 ft. \\
Power Consumption &  10 kW \\
  \hline
  \end{tabular}
  \label{table:perf}
  \end{table}

\begin{figure}[bt]
\begin{center}
\raisebox{1cm}{
\includegraphics[width=150mm]{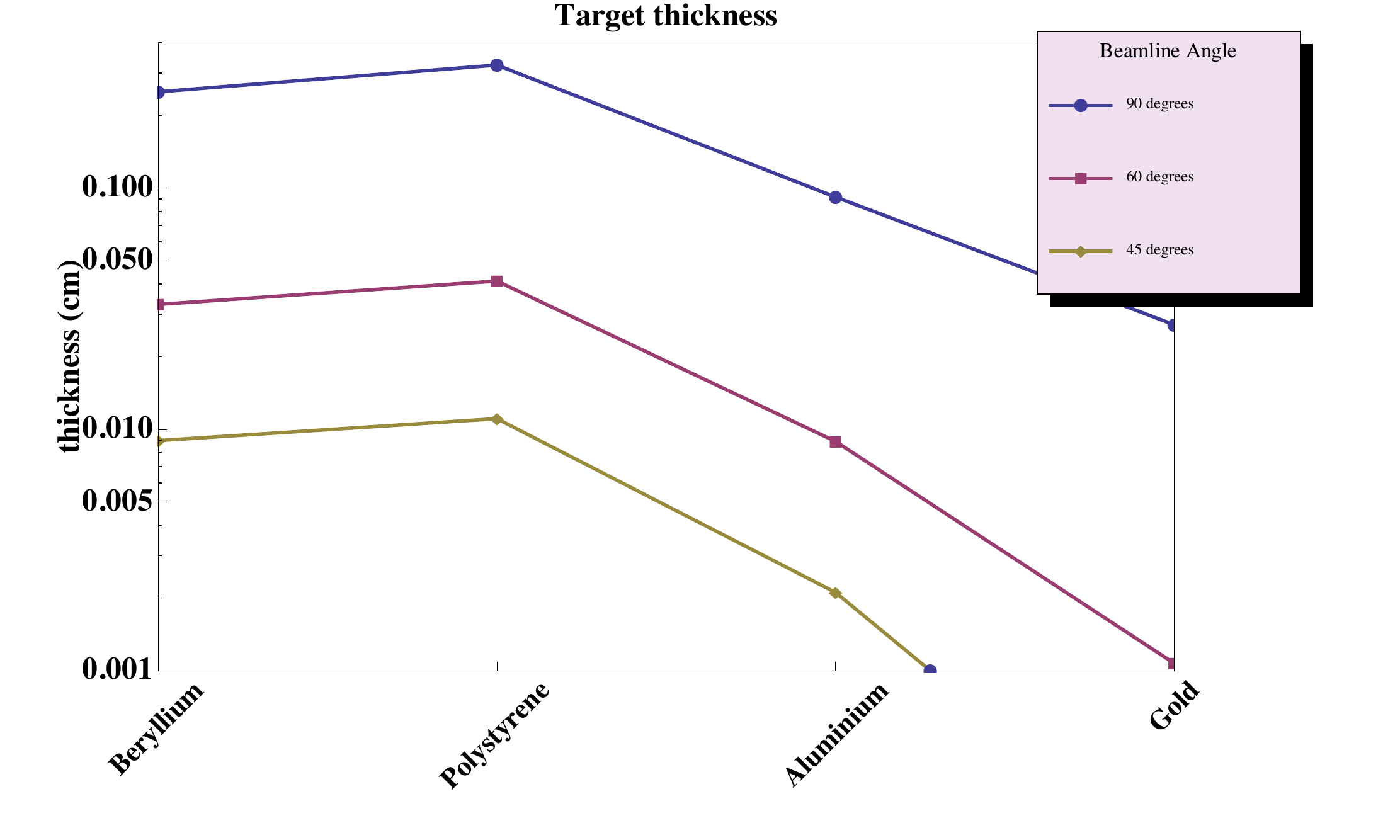}
}
\caption[Calculated thickness of scattering foils needed to produce 1 scattered electron/pulse in an aperture of $1^0$ or $10^{-3}$ steradians.]{\label{fig:foil}
Calculated thickness (in centimeters) of scattering foils needed to produce 0.5 scattered electron/pulse in an aperture of $1^0$ or $10^{-3}$ steradians.}
\end{center}
\end{figure}

\subsection{Scattered rates}

		We calculate the scattered electron rate starting from the Mott scattering formula (eqns. 1-5). The term $\propto Z\alpha _{EM}$ in eqn. 2 is often not included  
		but becomes significant above $Z\simeq6$. For all targets we consider the form factor suppression is significant, even at this
	low energy. Hofstadter's measurements were taken with beams in the 100 MeV energy range. It has been useful to compare this calculation with his
	cross section measurements- particularly Be-120 MeV, H and C-187 MeV, Au- 84 MeV.
	
		To parametrize the form factors we follow Hofstadter's original formula (eqn. 3-4). In this formula the charge density profile is fitted with an exponential form
		and the slope parameters are given in his papers (see Table 2). The slope parameter,a ,is defined by r=2 a$\sqrt{3}$, where r is the charge radius. Since the $A^{1/3}$ form doesn't fit the radii of lighter nuclei there isn't an obvious
		way to calculate it so we use his measured values.
		
		In this problem, since we require a scattered rate of 1/2 event/pulse, we invert the usual expression for the rate (i.e. rate=flux$\times\sigma\times\rho\times$t) to find the target thickness,t , needed to get this rate.

\begin{equation}
(\frac{d\sigma}{d\Omega})_{Rutherford}=1/4(Z\cdot\alpha _{EM})^2\frac{(\hbar c)^2}{E_e^2}Cosec(\theta/2)^4
\end{equation}

\begin{equation}
(\frac{d\sigma}{d\Omega})_{Mott}=(\frac{d\sigma}{d\Omega})_{Rutherford}\cdot Cos(\theta /2)^2(1+\frac{\pi Z \alpha _{EM} Sin(\theta /2)(1-Sin(\theta /2))}{Cos(\theta /2)^2})
\end{equation}

\begin{equation}
\rho(r)=\frac{1}{8\pi (a)^3}Exp(-r/a)
\end{equation}

\begin{equation}
FormFactor(Q)=\frac{4 \pi  \int _0^{\infty }r \rho (r,a) \sin (r Q)dr}{Q}
\end{equation}

\begin{equation}
(\frac{d\sigma}{d\Omega})_{Hofstadter}=(\frac{d\sigma}{d\Omega})_{Mott} FormFactor(Q)^2
\end{equation}

\begin{table}[ht]
\caption{Slope parameter}
\centering
\begin{tabular}{c c c}
\hline\hline
  & a(Fermi)& $r_0 A^{1/3}/(2\sqrt{3}$)  \\[0.5ex]
  \hline
 Beryllium & 0.64 & 0.780609 \\
 Carbon& 0.69 & 0.859171 \\
Gold & 2.3 & 2.18361 \\
  \hline
  \end{tabular}
  \label{table:a}
  \end{table}

	We obtain good agreement with available data in Ref.\cite{Hofstadter} at several energies and for H through Au targets.
This justifies the use of the exponential fit for the charge distribution.

\begin{figure}[bt]
\begin{center}
\raisebox{1cm}{
\includegraphics[width=75mm]{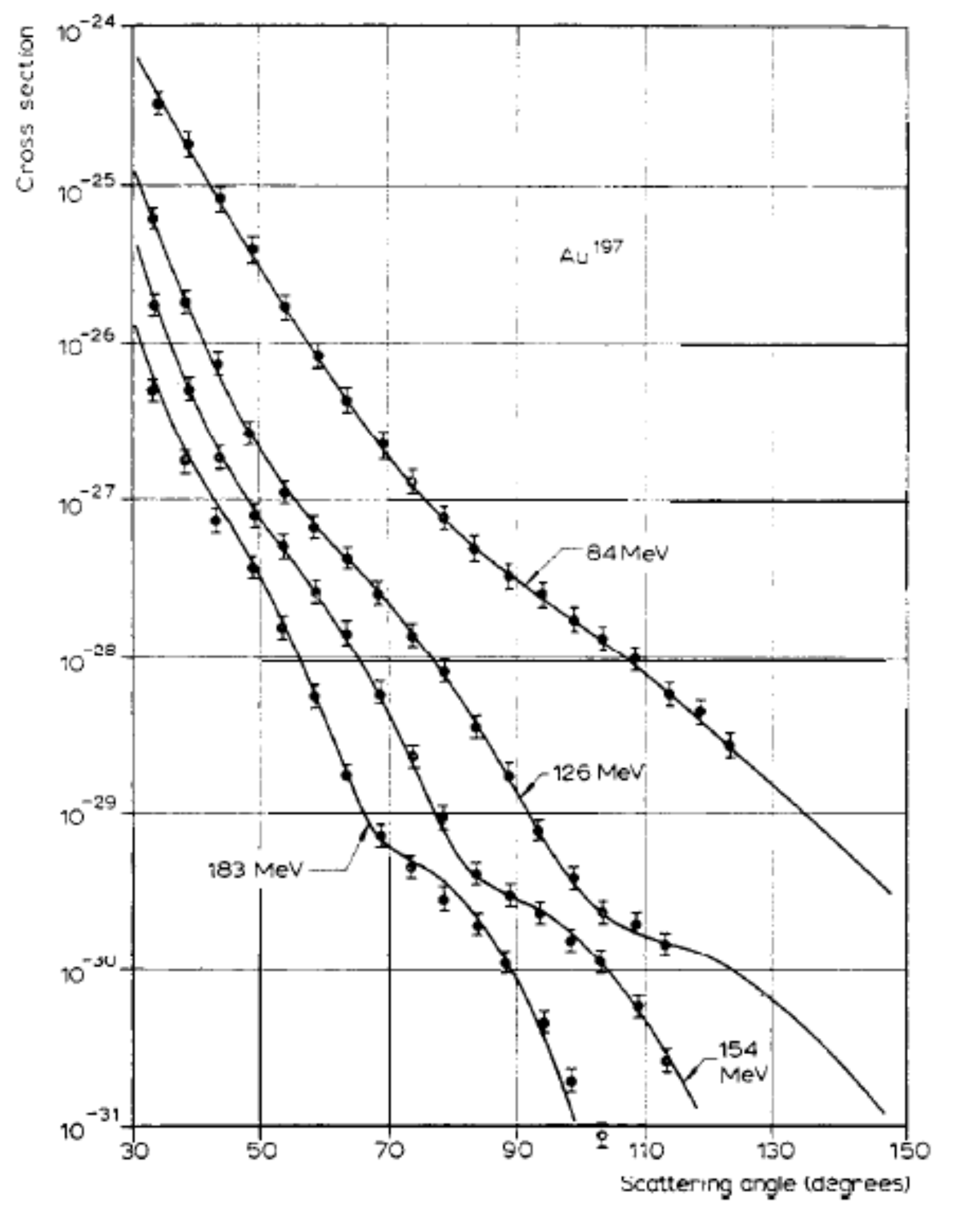}
\includegraphics[width=75mm]{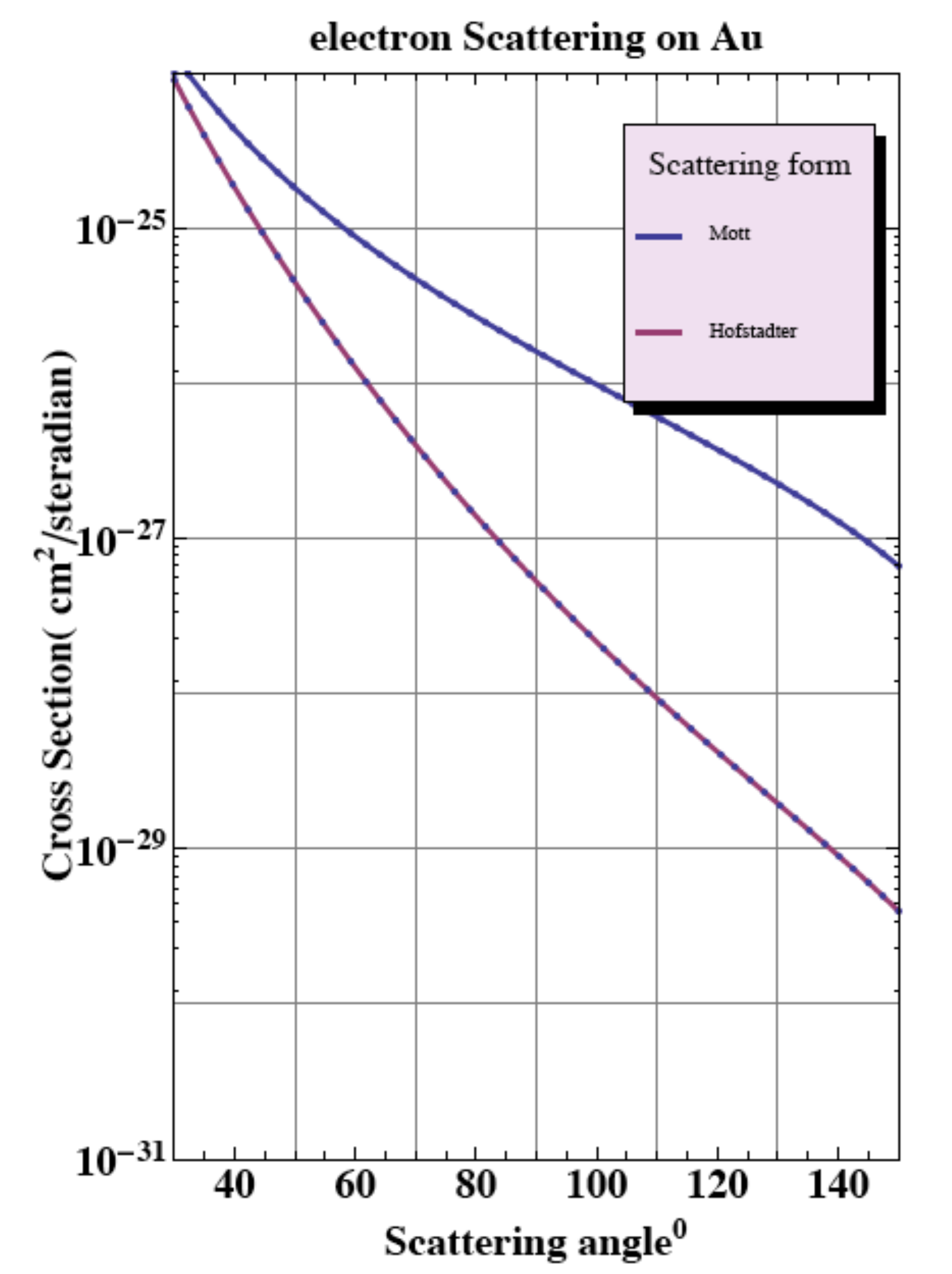}
}
\caption[ Elastic scattering scattering data on Au at 84 MeV (upper curve) from Ref. \cite{Hofstadter} compared with this calculation.]{\label{fig:Au_sig}
Elastic scattering data on Au at 84 MeV (upper curve) from Ref. \cite{Hofstadter}(left) compared with this calculation (right). Also shown in the upper curves of this calculation are the cross sections obtained without form factor suppression.}
\end{center}
\end{figure}

	In Fig.~\ref{fig:Au_sig}, for example, we show a comparison of this calculation with Hofstadter's data for gold. This calculation misses some of the diffractive fine structure seen with high energy beams but agrees
	everywhere to within a factor of 2.
	
	The results for target thickness are shown in Fig~\ref{fig:foil} and also in Table 3.

 \begin{table}[ht]
\caption{Calculated foil thicknesses (cm)}
\centering
\begin{tabular}{c c c c c}
\hline\hline
 & Beryllium& Polystyrene & Aluminum & Gold \\
 $45^0$ & 0.00898032 & 0.011074 & 0.00210775 & 0.000146681 \\
 $60^0$ & 0.0329445 & 0.0411883 & 0.00891158 & 0.00107188 \\
 $90^0$ & 0.250656 & 0.323623 & 0.0918123 & 0.0271518\\
  \hline
  \end{tabular}
  \label{table:a}
  \end{table}

\subsection{Beam emittance}

\begin{figure}[bt]
\begin{center}
\raisebox{1cm}{
\includegraphics[width=150mm]{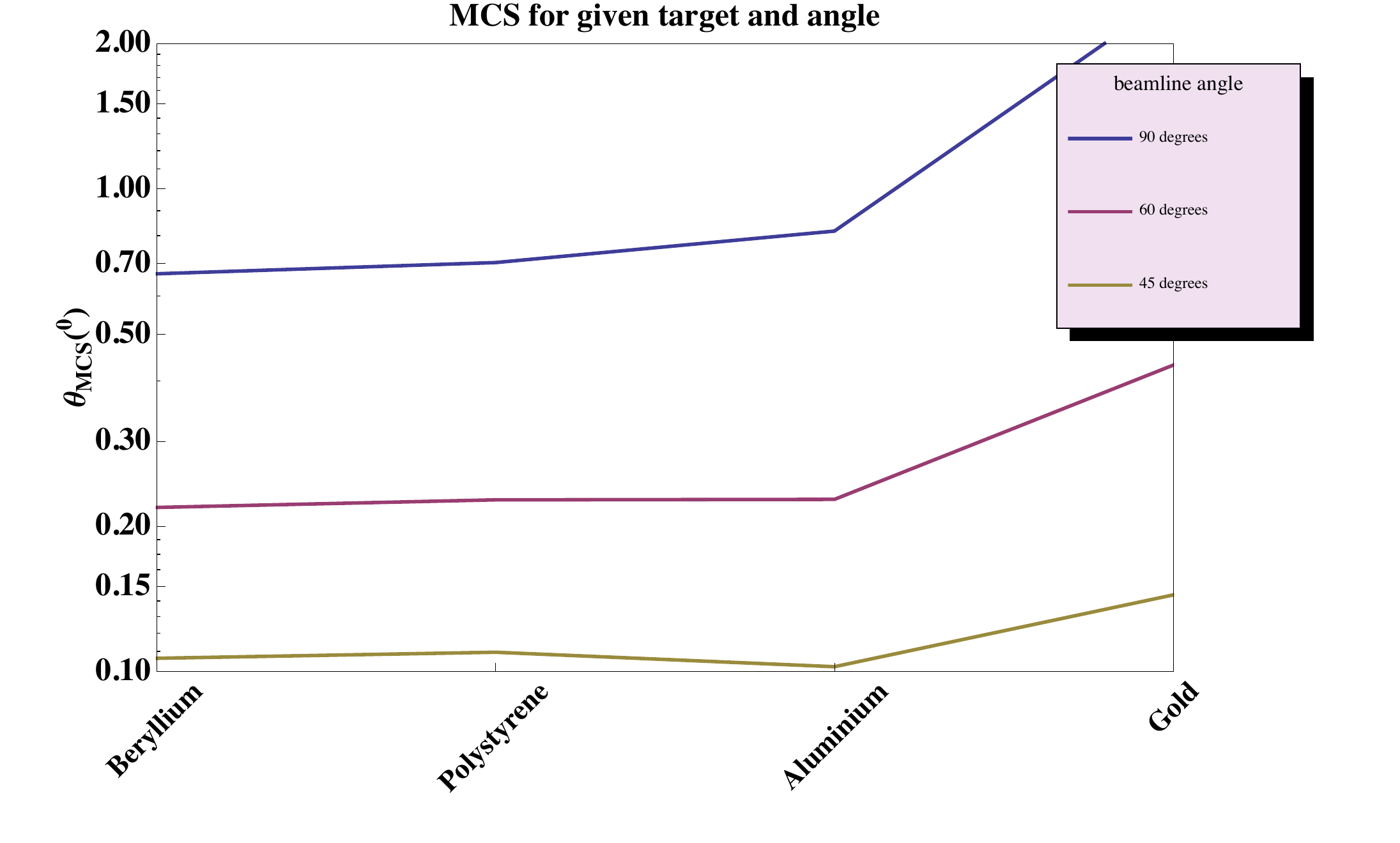}
}
\caption[Multiple Coulomb scattering in the forward beam due to different choices of foil thickness.
]{\label{fig:mcs}
Multiple Coulomb scattering in the forward beam due to different choices of foil thickness.}
\end{center}
\end{figure}

		For the beam to be useful downstream of the scattering target it would be best to keep the target thickness to a
	minimum and reduce multiple scattering in the target. The nonprojected multiple scattering distribution is characterized by $\theta_0$,
	which is the standard deviation of the gaussian approximation for the forward beam spread, and can be expressed as
	
\begin{equation}
\theta_0=  \frac{13.6 MeV}{E_e}\sqrt{\frac{t}{X_0}}\cdot(1 + .038\cdot Ln(\frac{t}{X_0})).
\end{equation}

	where t, the target thickness, and $X_0$, the radiation length of the material, are given in centimeters and the incident electron energy, $E_e$ is in MeV.
	Foils thick enough to produce a scattered beam at $90^0$ also lead to a significant emittance blowup ($\theta_{MCS}\sim.7^0$)  -see Fig~\ref{fig:mcs} and
	table 4 while, with a $45^0$ secondary beam, the multiple scattering angle in the primary beam is only $\sim 0.1^0$.
	
		Since we find for light targets that t/$X_0$ is smaller than $\alpha_{EM}$ the line shape of scattered electrons is dominated by inner Bremsstrahlung
		and therefore has negligible distortion (dN$_\gamma/dk\sim\alpha_{EM}/k$).

  \begin{table}[ht]
\caption{rms Multiple Coulomb Scattering Angle($^0$)}
\centering
\begin{tabular}{c c c c c}
\hline\hline
 & \text{Beryllium} & \text{Polystyrene} & \text{Aluminum} & \text{Gold} \\
  \hline
 $45^0$ & 0.106529 & 0.109651 & 0.102343 & 0.144157 \\
 $60^0$ & 0.21874 & 0.226824 & 0.227326 & 0.431394 \\
 $90^0$ & 0.666669 & 0.703342 & 0.817344 & 2.51229\\
 \hline
 \end{tabular}
 \label{table:a}
 \end{table}

\subsection{ Secondary peaks in the electron spectrum}

	One feature of this method is that for certain targets and scattering angles the electron spectrum has secondary peaks due
to energy loss through excitation of nuclear levels. If the detector can resolve these peaks which in $C^{12}$, for example, have a spacing of 4.4 MeV,
this would be a nice demonstration of performance. 

\begin{figure}[bt]
\begin{center}
\raisebox{1cm}{
\includegraphics[width=75mm]{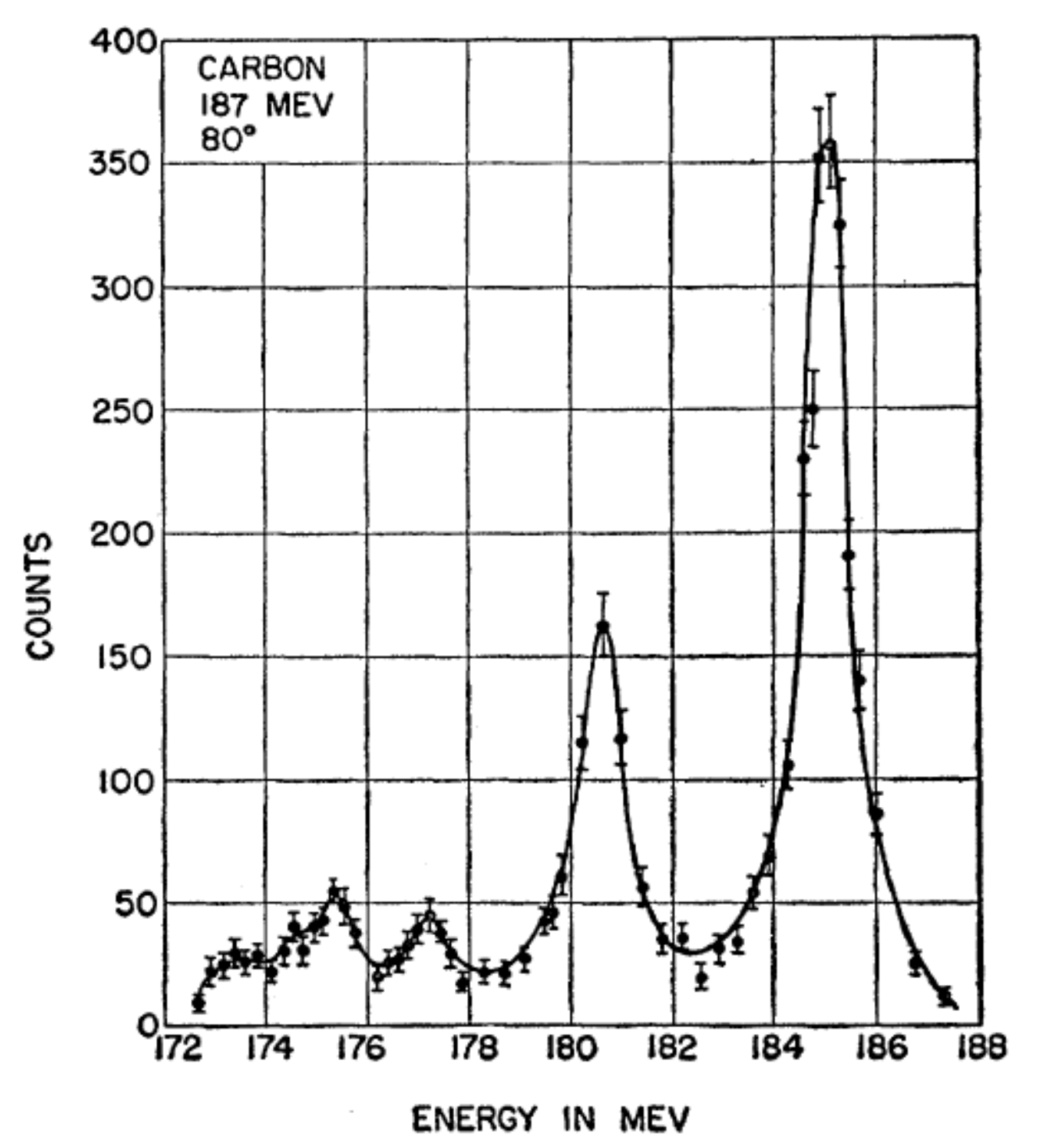}
\includegraphics[width=75mm]{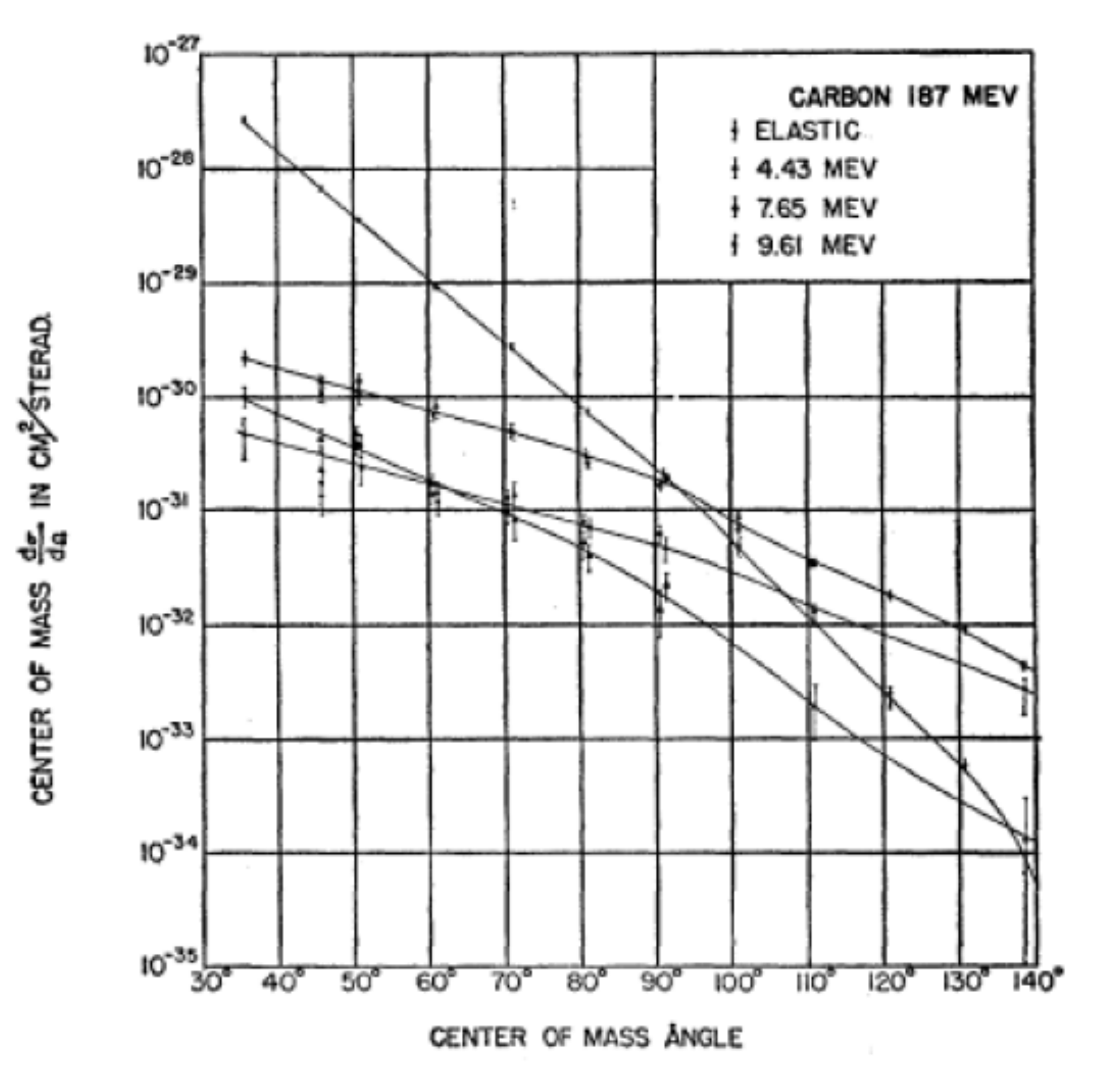}
}
\caption[  Electron scattering data on $C^{12}$ from Ref. \cite{Hofstadter}. The left panel shows the
energy spectrum seen at one angle ($80^0$. This spectrum is results from energy loss due to nuclear excitation since $E_{scattered}=E_{beam}-E_{excitation}$. The right panel shows the angular dependence of the elastic and inelastic rates.]{\label{fig:peaks}
The left panel shows the
energy spectrum seen at one angle ($80^0$). This spectrum results from energy loss due to nuclear excitation since $E_{scattered}=E_{beam}-E_{excitation}$. The right panel shows the angular dependence of the elastic and inelastic rates.}
\end{center}
\end{figure}

The case of $C^{12}$ is illustrated in Fig.~\ref{fig:peaks} where the left panel shows the
energy spectrum seen at an angle of $80^0$ degrees. The right panel shows the angular dependence of elastic and inelastic cross sections at 187 MeV. At roughly
90 degrees the 1st inelastic peak and the elastic peak become equal.

\subsection{Costs and Impact}

	We are looking at several approaches to costing this accelerator. There are
a number of similar accelerators which are soon de-commissioning and it is clear that 
many components will become available for free. Below we list the component costs based on
firm quotes in most cases. For example, the sections are from a SLAC design fabricated in China,
the photoinjector quotes are from AES and RadiaBeam and the laser cost is from a 2009 quote to
ATF.

	The engineering and maintenance costs of a custom built accelerator could be significant.
For example, assembly of the 100 MeV machine given in Table 5 would certainly require a year of work by
an RF/laser engineer.

	Another approach is to order a turn-key machine from one of 2 companies-AES on Long Island or RadiaBeam in Santa Monica,Ca. 
This solution could be completed in $\sim18$ months and would cost $\sim 50\%$ more than the component cost given below\cite{RadiaBeam}. 

	 Maintenance costs would depend a lot on the running scenario. Although such machines run at
essentially 100$\%$ duty cycle, doing so assumes that there is a staff on hand to operate the machine- at least
a technician and a laser engineer. If a lower duty cycle is needed and
one can absorb unanticipated down times then you wouldn't need more than 1 person on staff.

		Generally the impact on the infrastructure is the 10 kWatt continuous power required. This machine would have a $\sim$ 600 $ft^2$ footprint and would be at least ~30 ft long. For example, it could be 30ft x 20ft or 60ft x 10ft. For installation,
the accelerator could be broken down into smaller modules with dimensions 10x6x6 $ft^3$ (in any
orientation) and lowered into DUSEL.

 \begin{table}[ht]
\caption{rough Costs}
\centering
\begin{tabular}{c c }
\hline\hline
Component costs &   \\
  \hline
 TiSaphire Laser (ATF quote)&400k \\
   Photoinjector (AES, Radia Beam)&350k \\
  2 Klystrons &250k \\
  2 Modulators  & 500k\\
  2 Sections  & 300k \\
  Low level RF, etc.  &200k \\
 Supports, etc. &100k\\
 \hline
Total & 2.1M USD \\
\hline
\hline
   \end{tabular}
  \label{table:a}
  \end{table}

\end{document}